\documentclass[11pt,fleqn]{article}

\usepackage{comment}

\usepackage{amsmath,amssymb,graphicx,mdframed,pslatex,algorithm2e,verbatim,wrapfig}

\includecomment{book}

\usepackage{acronym}

\acrodef{API}{Application Programmer Interface}
\acrodef{BFS}{Breadth-First Search}
\acrodef{BSP}{Bulk Synchronous Parallelism}
\acrodef{CAF}{Co-Array Fortran}
\acrodef{CG}{Conjugate Gradients}
\acrodef{CREW}{Concurrent Read, Exclusive Write}
\acrodef{CSP}{Communication Sequential Processes}
\acrodef{DAG}{Directed Acyclic Graph}
\acrodef{DSL}{Domain-Specific Language}
\acrodef{DTD}{Distributed Termination Detection}
\acrodef{FDM}{Finite Difference Method}
\acrodef{FEM}{Finite Element Method}
\acrodef{FMM}{Fast Multipole Method}
\acrodef{FSM}{Finite State Machine}
\acrodef{GA}{Global Arrays}
\acrodef{GPU}{Graphics Processing Unit}
\acrodef{GS}{Gauss-Seidel}
\acrodef{HPC}{High Performance Computing}
\acrodef{HPF}{High Performance Fortran}
\acrodef{IMP}{Integrative Model for Parallelism}
\acrodef{IR}{Intermediate Representation}
\acrodef{ISA}{Incremental Single Assignment}
\acrodef{MPI}{Message Passing Interface}
\acrodef{NUMA}{Non-Uniform Memory Access}
\acrodef{OO}{Object-Oriented}
\acrodef{PDE}{Partial Differential Equation}
\acrodef{PGAS}{Partitioned Global Address Space}
\acrodef{PRAM}{Parallel Random Access Machine}
\acrodef{RPC}{Remote Procedure Call}
\acrodef{SMP}{Symmetric Multi-Processor}
\acrodef{SSA}{Static Single Assignment}
\acrodef{SPMD}{Single Program Multiple Data}
\acrodef{SSSP}{Single-Source Shortest Path}
\acrodef{UPC}{Universal Parallel C}

\usepackage{etoolbox}

\usepackage[pdftex,colorlinks]{hyperref}
\usepackage{xr-hyper}
\hypersetup{bookmarksopen=true}
\def\heading#1{\paragraph*{\textbf{#1}\kern1em}\ignorespaces}

\expandafter\ifx\csname corollary\endcsname\relax
    \newtheorem{corollary}{Corollary}
\fi
\expandafter\ifx\csname lemma\endcsname\relax
    
\fi
\expandafter\ifx\csname theorem\endcsname\relax
    
\fi
\expandafter\ifx\csname proof\endcsname\relax
 \newenvironment{proof}{\begin{quotation}\small\sl\noindent Proof.\ \ignorespaces}
     {\end{quotation}}
\fi
\expandafter\ifx\csname definition\endcsname\relax
  \newtheorem{definition}{Definition}
\fi
\expandafter\ifx\csname example\endcsname\relax
  
\fi
\expandafter\ifx\csname remark\endcsname\relax
  \newtheorem{remark}{Remark}
\fi

\def\kw#1{\mathord{\mathrm{#1}}}%{\ifmmode \mathord{\mathrm{#1}} \else $\mathord{\mathrm{#1}}$ \fi}

\usepackage{framed}

\usepackage{mdframed}

\def\n{\bgroup\tt\catcode`\_=12 \let\next=}

\def\twocode {\afterassignment\twocodeb\def\nexta}
\def\twocodeb{\bgroup \catcode`\_=12\relax
              \afterassignment\twocodec\global\def\nextb}
\def\twocodec{\egroup %\show\nexta \show\nextb
  \par\smallskip
  \hskip\unitindent $\vcenter{\hsize=.37\hsize$\nexta$}\quad
   \vcenter{\hsize=.5\hsize\footnotesize\tt\nextb}$
  \par\smallskip
}

\def\n{\bgroup\tt\catcode`\_=12 \let\next=}

\title{A mathematical formalization of data parallel operations}
\author{Victor Eijkhout\thanks{{\tt
      eijkhout@tacc.utexas.edu}, Texas Advanced Computing Center, The
    University of Texas at Austin}}

\begin{document}
\maketitle

\begin{abstract}
  We give a mathematical formalization of `generalized data parallel'
  operations, a concept that covers such common scientific kernels as
  matrix-vector multiplication, multi-grid coarsening, load
  distribution, and many more.
  We show that from a compact specification such computational aspects
  as MPI messages or task dependencies can be automatically derived.
\end{abstract}

\acresetall

\section{Introduction}

In this paper we give a rigorous formalization of several scientific
computing concept that are commonly used, but rarely defined;
specifically distributions,
data parallel operations, and `halo regions'.
Taken together, these concepts allow a minimal specification of an
algorithm by the programmer to be translated into the communication
and synchronization constructs that are usually explicitly programmed.

Looking at it another way, we note that communication and
synchronization in a parallel code stem from both algorithm and data
distribution properties. The contribution of this work is then that we
have found a separation of concerns that allows the programmer to
specify them separately, while the resulting communication and
synchronization is derived formally and therefore automatically.

We start with a motivating example in
section~\ref{sec:threepoint-example}, followed by a formal derivation
in section~\ref{sec:formal}. We conclude by discussing the
practical ramifications of our work.

\section{Motivating example}
\label{sec:threepoint-example}
% -*- latex -*-
%%%%%%%%%%%%%%%%%%%%%%%%%%%%%%%%%%%%%%%%%%%%%%%%%%%%%%%%%%%%%%%%
%%%%%%%%%%%%%%%%%%%%%%%%%%%%%%%%%%%%%%%%%%%%%%%%%%%%%%%%%%%%%%%%
%%%%
%%%% This text file is part of the theory writeup on the
%%%% Integrative Model for Parallelism,
%%%% copyright Victor Eijkhout (eijkhout@tacc.utexas.edu) 2014-6
%%%%
%%%% threepoint.tex : threepoint difference example as introduction
%%%%     of alpha/beta/gamma
%%%%
%%%%%%%%%%%%%%%%%%%%%%%%%%%%%%%%%%%%%%%%%%%%%%%%%%%%%%%%%%%%%%%%
%%%%%%%%%%%%%%%%%%%%%%%%%%%%%%%%%%%%%%%%%%%%%%%%%%%%%%%%%%%%%%%%

We consider a simple data parallel example, and show how it leads
to the basic distribution concepts of \ac{IMP}: the three-point operation
\[ \forall_i\colon y_i=f(x_i,x_{i-1},x_{i+1}) \]
which describes for instance the 1D heat equation
\[ y_i = 2x_i-x_{i-1}-x_{i+1}. \]
(Stencil operations are much studied; see e.g.,~\cite{Tang:2011:pochoir}
and the polyhedral model, e.g.,~\cite{Dathathri:generating-movement}.
However, we claim far greater generality for our model.)
We  illustrate this graphically by depicting the input and output vectors,
stored distributed over the processors by contiguous blocks,
and the three-point combining operation:

\includegraphics[scale=.12]{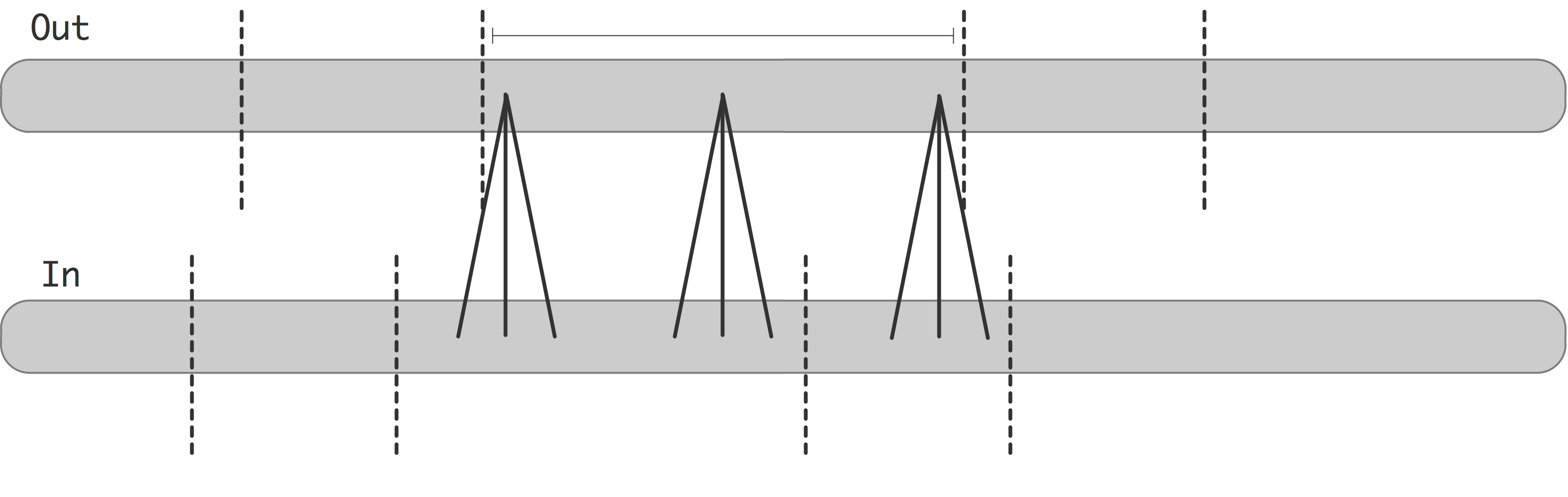}

The distribution indicated by vertical dotted lines
we call the $\alpha$-distribution for the input,
and the $\gamma$-distribution for the output.
These distributions are mathematically given as
an assignment from processors to sets of indices:
\[ \alpha\colon p\mapsto [ i_{p,\min},\ldots,i_{p,\max}]. \]
The traditional concept of distributions in parallel programming systems
is that of an assignment of data indices to a processor,
reflecting that each index `lives on' one processor,
or that that processor is responsible for computing that index of the output.
We turn this upside down: we define a distribution as a mapping from 
processors to indices. This means that an index can `belong' to more than
one processor. (The utility of this for redundant computing is
obvious. However, it will also seen to be crucial for our general framework.)

For purposes of exposition we will now equate
the input $\alpha$-distribution and the output $\gamma$-distribution,
although that will
not be necessary in general.

\includegraphics[scale=.12]{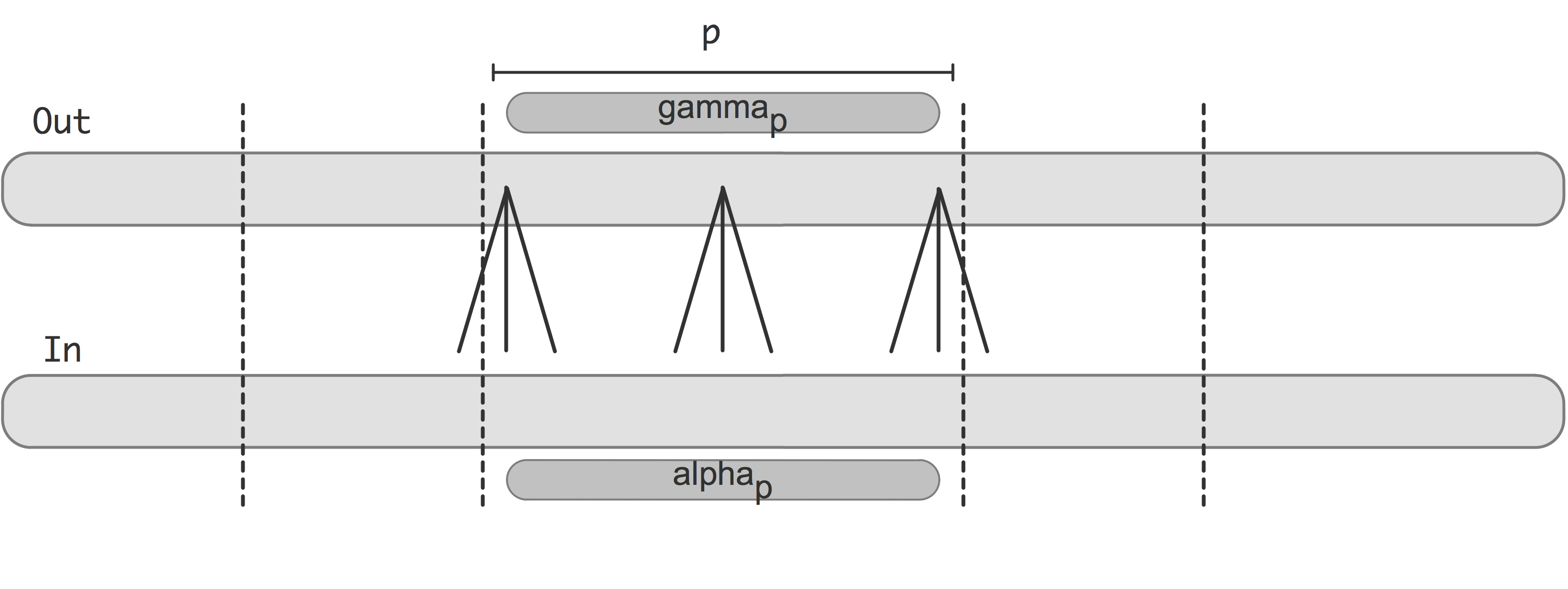}

This 
picture shows how, for the three-point operation,
some of the output elements on processor~$p$
need inputs that are not present on~$p$.
For instance, the computation of~$y_i$
for $i_{p,\min}$ takes an element from processor~$p-\nobreak1$.
This gives rise to what we call the $\beta$-distribution:
\begin{quotation}
\begin{mdframed}{$\beta(p)$~is the set
of indices that processor~$p$ needs to compute the indices in~$\gamma(p)$.}
\end{mdframed}
\end{quotation}

The next illustration depicts the different distributions
for one particular process:

\includegraphics[scale=.12]{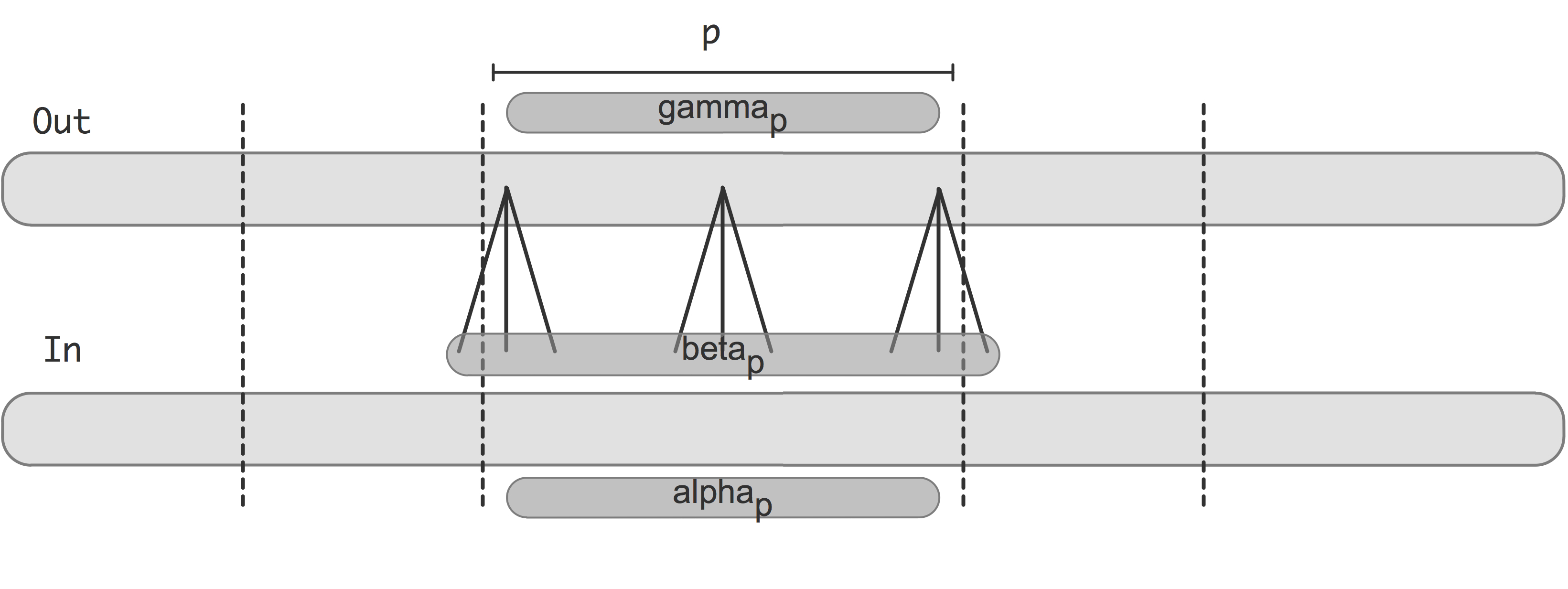}

Observe that the $\beta$-distribution, unlike the $\alpha$
and~$\gamma$ ones, is not disjoint: certain elements live on
more than one processing element. It is also, unlike
the $\alpha$ and $\gamma$ distributions, not specified by
the programmer: it is derived from the $\gamma$-distribution
by applying the shift operations of the stencil. That is,
\begin{quotation}
  \begin{mdframed}{The $\beta$-distribution brings together properties of the algorithm
    and of the data distribution.}
  \end{mdframed}
\end{quotation}
We will formalize
this derivation below.

\subsection{Definition of parallel computing}
\label{sec:3pt-dag}

This gives us all the ingredients for reasoning about parallelism.
Defining a `kernel' as a mapping from one distributed data set
to another, and a `task' as a kernel on one particular process(or),
all data dependence of a task results from transforming
data from $\alpha$ to $\beta$-distribution.
By analyzing the relation between these two we derive at dependencies
between processors or tasks: each processor~$p$ depends on
some predecessors~$q_i$, and this set of predecessors can be derived
from the $\alpha,\beta$ distributions: $q_i$~is a predecessor if
\[ \alpha(q_i)\cap\beta(p)\not=\emptyset. \]

\includegraphics[scale=.12]{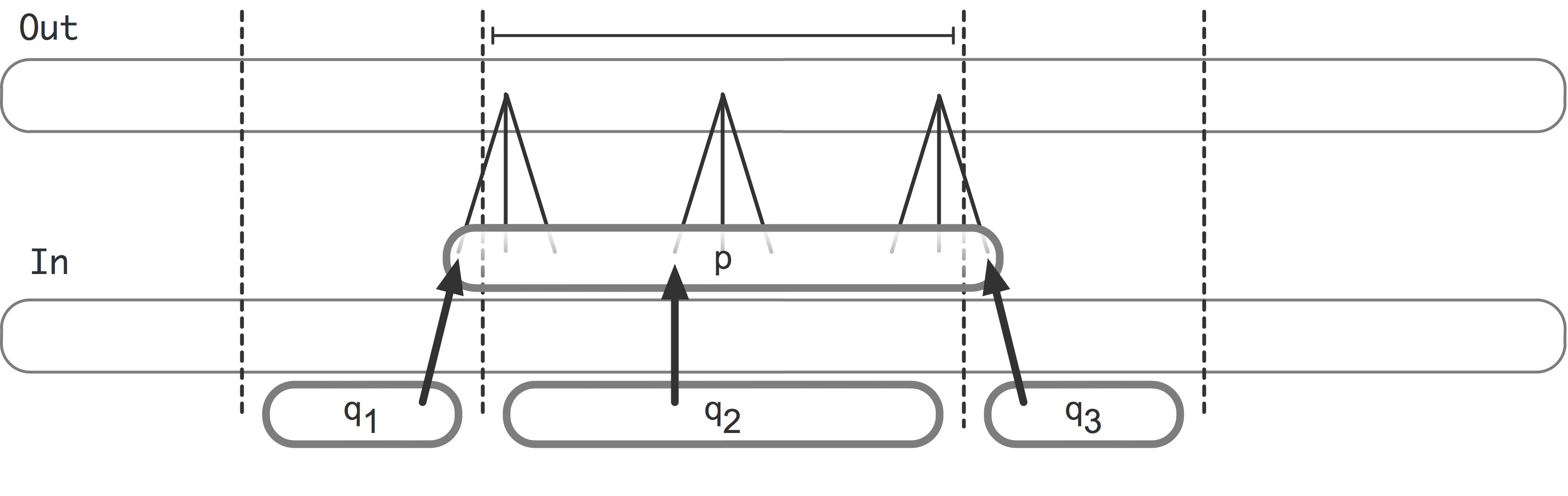}

In message passing, these dependences obviously corresponds
to actual messages: for each process~$p$, the processes~$q$
that have elements in $\beta(p)$
send data to~$p$. (If $p=q$, of course at most a copy is called for.)
Interestingly, this story has an interpretation in tasks on shared
memory too.  If we identify the $\alpha$-distribution on the input
with tasks that produce this input, then the $\beta$-distribution
describes what input-producing tasks a task~$p$ is dependent on. In
this case, the transformation from $\alpha$ to $\beta$-distribution
gives rise to a \emph{dataflow} formulation of the
algorithm.

\subsection{Programming the model}
% -*- latex -*-
%%%%%%%%%%%%%%%%%%%%%%%%%%%%%%%%%%%%%%%%%%%%%%%%%%%%%%%%%%%%%%%%
%%%%%%%%%%%%%%%%%%%%%%%%%%%%%%%%%%%%%%%%%%%%%%%%%%%%%%%%%%%%%%%%
%%%%
%%%% This text file is part of the theory writeup on the
%%%% Integrative Model for Parallelism,
%%%% copyright Victor Eijkhout (eijkhout@tacc.utexas.edu) 2014-6
%%%%
%%%% programthreepoint.tex : IMP programming based on threepoint example
%%%%
%%%%%%%%%%%%%%%%%%%%%%%%%%%%%%%%%%%%%%%%%%%%%%%%%%%%%%%%%%%%%%%%
%%%%%%%%%%%%%%%%%%%%%%%%%%%%%%%%%%%%%%%%%%%%%%%%%%%%%%%%%%%%%%%%

\label{sec:define-signature}

In our motivating example we showed how the concept of
`$\beta$-distribution' arises, and the role it plays combining
properties of the data distributions and of the algorithm's data dependencies.
This distribution generalizes concepts such as the `halo region'
in distributed stencil calculations, but its applicability extends to
all of (scientific) parallel computing.
For instance, for collectives we can define a $\beta$-distribution,
which is seen to equal the $\gamma$-distribution.

It remains to be argued that the $\beta$ distribution can actually
be used as the basis for a software system.
To show this, we associate with the function~$f$
that we are computing an expression of the algorithm
(not the parallel!) data dependencies,
called the `signature function',
denoted~$\sigma_f$. For instance for the computation
of $y_i=f(x_i,x_{i-1},x_{i+1})$, the signature function is
\[ \sigma_f(i)=\{i,i-1,i+1\}. \]
With this, we state 
(without proof; for which see section~\ref{sec:beta-theorem} and~\cite{IMP-01})
that
\[ \beta=\sigma_f(\gamma). \]
If follows, if the programmer can specify the data dependencies of
the algorithm, a~compiler/runtime system can derive the $\beta$ distribution,
and from it, task dependencies and messages for parallel execution.

Specifying the signature function is quite feasible, but the precise
implementation depends on the context. For instance, for regular applications
we can adopt a syntax similar to stencil compilers such as
the Pochoir compiler~\cite{Tang:2011:pochoir}. For sparse matrix
applications the signature function is isomorphic to the adjacency
graph; for collective operations, $\beta=\gamma$ often holds; et cetera.

\section{Formal definition}
\label{sec:formal}
% -*- latex -*-
%%%%%%%%%%%%%%%%%%%%%%%%%%%%%%%%%%%%%%%%%%%%%%%%%%%%%%%%%%%%%%%%
%%%%%%%%%%%%%%%%%%%%%%%%%%%%%%%%%%%%%%%%%%%%%%%%%%%%%%%%%%%%%%%%
%%%%
%%%% This text file is part of the theory writeup on the
%%%% Integrative Model for Parallelism,
%%%% copyright Victor Eijkhout (eijkhout@tacc.utexas.edu) 2014-6
%%%%
%%%% distributionmath.tex : mathematical definitions of distributions.
%%%%
%%%%%%%%%%%%%%%%%%%%%%%%%%%%%%%%%%%%%%%%%%%%%%%%%%%%%%%%%%%%%%%%
%%%%%%%%%%%%%%%%%%%%%%%%%%%%%%%%%%%%%%%%%%%%%%%%%%%%%%%%%%%%%%%%

\subsection{Data parallel computation}

The \acf{IMP} is a theory of data parallel functions. By this we mean
functions where each element of a distributed output object is computed from
one or more elements of one or more distributed input
objects.
\begin{itemize}
\item Without loss of generality we limit ourselves to a single
  input object.
\item Since all output elements can be computed independently of each
  other, we call this a `data parallel' function. In our context this
  has no connotations of SIMD or synchronization; it merely expresses
  independence.
\end{itemize}
Formally, a data parallel computation is the use of a 
function with a single output to compute the elements of a distributed object:
\[ \kw{Func} \equiv  \kw{Real}^k\rightarrow \kw{Real} \]
where $k$ is some integer.

Since we will mostly talk about indices rather than data, we define
$\kw{Ind} \equiv 2^N$
and we describe the structure of the data parallel computation
through a `signature function':
\[ \kw{Signature}\equiv N\rightarrow \kw{Ind}. \]

In our motivating example, where we computed
$y_i=f(x_{i-1},x_i,x_{i+1})$, our signature function was
\[ \sigma_f\equiv i\mapsto \{ i-1,i,i+1 \}. \]

\begin{itemize}
\item The signature function can be compactly rendered in cases of a
  stencil computation.
\item In general it describeds the bi-partite graph of data
  dependencies. Thus, for sparse computations it is isomorphic to the
  sparse matrix, and can be specified as such.
\item In certain cases, the signature function can be most compactly
  be rendered as a function recipe. For instance, for 1D multigrid
  restriction it would be given as $\sigma(i)=\{2i,2i+1\}$.
\item For collectives such as an `allreduce', the signature function
  expresses that the output is a function of all inputs:
  $\forall_i\colon\sigma(i)=N$.
\end{itemize}

\subsection{Distributions}

We now formally define distributions as mappings from processors to
sets of indices:
\[ \kw{Distr}\equiv \kw{Proc} \rightarrow \kw{Ind}. \]
Traditionally, distributions are considered as mappings from data elements 
to processors, which assumes a model where a data element lives uniquely 
on one processor. By turning this definition around we have an elegant way of describing:
\begin{itemize}
\item Overlapping distributions such as halo data, where data has been
  gathered on a processor for local operations. Traditionally, this is
  considered a copy of data `owned' by another processor.
\item Rootless collectives: rather than stating that all processors
  receive an identical copy of a result, we consider them to actually
  own the same item.
\item Redundant execution. There are various reasons for having
  operation executed redundantly on more than one processor. This can
  for instance happen in the top levels of a coarsening multilevel
  method, or in redundant computation
  for resilience. 
\end{itemize}

We now bring together the concepts of signature function and distribution:
\begin{enumerate}
\item We can extend the signature function concept, defined above as
  mapping integers to sets of integers, to a mapping from sets to sets:
%
% \[ \kw{Signature} \equiv \kw{Ind}\rightarrow \kw{Ind} \]
%
with the obvious definion that, for $\sigma\in\kw{Signature},S\in\kw{Ind}$:
\[ \sigma(S) = \{ \sigma(i)\colon i\in S \}. \]
In our motivating example,
\[ \sigma\bigl([i_{\min},i_{\max}]\bigr) = [i_{\min}-1,i_{\max}+1]. \]
\item We then extend this to distributions
%
% \[ \kw{Signature} \equiv \kw{Distr} \rightarrow \kw{Distr} \]
%
with the definition that
for $\sigma\in\kw{Signature}$ and $u\in\kw{Distr}$
\[ \sigma(u) \equiv p\mapsto \sigma(u(p)) \quad\hbox{where}\quad
   \sigma(u(p)) = \{ \sigma(i)\mid i\in u(p) \}
\]
\end{enumerate}

We now have the tools for our grand result.

\subsection{Definition and use of $\beta$-distribution}
\label{sec:beta-theorem}

Consider a data parallel operation $y=f(x)$ where $y$ has
distribution~$\gamma$, and $x$ has distribution~$\alpha$. We call a
local operation to be one where every processor has all
the elements of~$x$ needed to compute its part of~$y$. By the above
overloading mechanism, we find that the total needed input on
processor~$p$ is $\sigma\bigl(\gamma(p)\bigr)$.

This leads us to define a \emph{local operation} formally as:
\begin{definition}
  We call a kernel $y=f(x)$ a local operation if $x$~has
  distribution~$\alpha$, $y$~has distribution~$\gamma$, and
  \[ \alpha\supset \sigma_f(\gamma). \]
\end{definition}

That is, for a local operation every processor already owns all the
elements it needs for its part of the computation.

Next, we call $\sigma_f(\gamma)$ the `$\beta$-distribution' of a function~$f$:

\begin{definition}
  If $\gamma$ is the output distribution of a computation~$f$,
  we define
  the $\beta$-distribution as \[ \beta=\sigma_f(\gamma). \]
\end{definition}

Clearly, if $\alpha\supset\beta$, each processor has all its needed
inputs, and the computation can proceed locally. However, this is
often not the case, and considering the difference between $\beta$
and~$\alpha$ gives us the description of the
task/process communication:
\begin{corollary}
  If $\alpha$ is the input distribution of a data parallel operation,
  and $\beta$ as above, then processor~$q$ is a predecessor of
  processor~$p$ if \[ \alpha(q)\cap\beta(p)\not=\emptyset. \]
\end{corollary}
\begin{proof}
  The set $\beta(p)$ describes all indices needed by processor~$p$;
  if the stored elements in~$q$ overlap with this, the computation
  on~$q$ that produces these is a predecessor of the subsequent
  computation on~$p$.
\end{proof}

This predecessor relation takes a specific form depending on the
parallelism mode. For instance, in message passing it takes form of an
actual message from $q$ to~$p$; in a \ac{DAG} model such as OpenMP
tasking it becomes a `task wait' operation.

\begin{remark}
  In the context of \ac{PDE} based applications, our
  $\beta$-distribution corresponds loosely to the `halo' region. The
  process of constructing the $\beta$-distribution is implicitly part of
  such packages as PETSc~\cite{GrSm:petsc}, where the communication
  resulting from it is constructed in the
  \n{MatAssembly} call. Our work takes this ad-hoc calculation, and
  shows that it can formally be seen to underlie a large part of
  scientific parallel computing.
\end{remark}

\section{Practical importance of this theory}

The above discussion considered operations that can be described as
`generalized data parallel'. From such operations one can construct
many scientific algorithms. For instance, in a multigrid method a
red-black smoother is data parallel, as are the restriction and
prolongation operators.

In the \ac{IMP} model these are termed `kernels', and each kernel
gives rise to one layer of task dependencies; see
section~\ref{sec:3pt-dag}.
Taking together the dependencies for the single kernels
then gives us a complete task graph for a parallel execution;
the edges in this graph can be interpreted as MPI messages
or strict dependencies in a \ac{DAG} execution model.

Demonstration software along these lines has been built, showing
performance comparable to hand-coded software; see~\cite{IMP-19}.

\section{Summary}

In this paper we have given a rigorous mathematical definition of data
distributions and the signature function of a data parallel
operation. Our notion of data distribution differs from the usual
interpretation in that we map processors to data, rather than
reverse. The signature function appears implicitly in the literature, for
instance in stencil languages, but our explicit formalization seems
new.

These two concepts effect a separation of concerns in the description
of a parallel algorithm:
the data distribution is an expression of the parallel aspects of,
while the signature function is strictly a description of the
algorithm. The surprising result is that these two give rise to a
concept we define as the `$\beta$-distribution'; it can be derived
from data distribution and signature function, and it contains enough
information to derive the communication~/ synchronization aspects of
the parallel algorithm.

Demonstrating the feasibility of programming along these lines, we
mention our \acf{IMP} system, which implements these ideas, and is
able to perform competitively with traditionally coded parallel applications.

\section*{Acknowledgement}

This work was supported by NSF EAGER grant 1451204.
The code for this project is available at \url{https://bitbucket.org/VictorEijkhout/imp-demo}.

\bibliography{vle}

\begin{thebibliography}{1}

\bibitem{Dathathri:generating-movement}
Roshan Dathathri, Chandan Reddy, Thejas Ramashekar, and Uday Bondhugula.
\newblock Generating efficient data movement code for heterogeneous
  architectures with distributed-memory.
\newblock In {\em Proceedings of the 22nd international conference on Parallel
  architectures and compilation techniques}, PACT '13, pages 375--386,
  Piscataway, NJ, USA, 2013. IEEE Press.

\bibitem{IMP-01}
Victor Eijkhout.
\newblock {IMP} distribution theory.
\newblock Technical Report IMP-01, Integrative Programming Lab, Texas Advanced
  Computing Center, The University of Texas at Austin, 2014.

\bibitem{IMP-19}
Victor Eijkhout.
\newblock Report on {NSF EAGER} 1451204.
\newblock Technical Report IMP-19, Integrative Programming Lab, Texas Advanced
  Computing Center, The University of Texas at Austin, 2014.

\bibitem{GrSm:petsc}
W.~D. Gropp and B.~F. Smith.
\newblock Scalable, extensible, and portable numerical libraries.
\newblock In {\em Proceedings of the Scalable Parallel Libraries Conference,
  IEEE 1994}, pages 87--93.

\bibitem{Tang:2011:pochoir}
Yuan Tang, Rezaul~Alam Chowdhury, Bradley~C. Kuszmaul, Chi-Keung Luk, and
  Charles~E. Leiserson.
\newblock The pochoir stencil compiler.
\newblock In {\em Proceedings of the 23rd ACM symposium on Parallelism in
  algorithms and architectures}, SPAA '11, pages 117--128, New York, NY, USA,
  2011. ACM.

\end{thebibliography}
\bibliographystyle{plain}

\end{document}